\documentclass[prc,aps]{revtex4}
\usepackage{graphicx}
\usepackage{natbib}

\begin{document}

\title[The anatomy of atomic nuclei]{The anatomy of atomic nuclei: illuminating many-body wave functions through group-theoretical decomposition}

\author{Calvin W. Johnson}
\affiliation{San Diego State University, 5500 Campanile Drive, San Diego, California 92182, USA}


\begin{abstract}
With modern computers we can compute nuclear many-body wave functions with an astounding number of component, 
 $ > 10^{10}$.  But, aside from reproducing and/or predicting experiments, what do we learn from 
vectors with tens of billions of components?  One way to characterize wavefunctions is through irreducible representations of groups. 
I discuss briefly the history of group-theoretical characterization of nuclear wavefunctions, with an emphasis of using Lanczos-type methods to 
efficiently dissect arbitrary wavefunctions into group irreps. Although the resulting decompositions are often 
 fragmented over many irreps, one nonetheless finds powerful patterns. First, group decompositions along rotational bands
show coherent commonalities, supporting the picture of a shared ``intrinsic shape;'' this is also called \textit{quasi-dynamical symmetry}. 
Second, group decompositions for wave functions using 
both phenomenological and \textit{ab initio} forces are often very similar, despite vastly different origins and dimensionalities. Both of these results 
suggest a group theoretical decomposition can provide a robust ``anatomy'' of many nuclear wave functions. This in turn supports the idea of 
using symmetry-based many-body frameworks for calculations.

\end{abstract}

\maketitle

\section{The blessing and the curse of large-scale computing}

The quantum theory of atomic nuclei is challenging.  The force between nucleons is strong, short-ranged, and complicated \cite{machleidt1987bonn}. 
For many decades nuclei were primarily modeled through 
phenomenological descriptions \cite{br88}. 
This changed starting two decades ago, thanks to the confluence of several factors: the development of forces which describe nucleon-nucleon scattering to 
high precision \cite{stoks1994construction,wiringa1995accurate}, 
especially the more recent introduction of chiral effective field theory \cite{van1999effective,entem2003accurate,epelbaum2006few,machleidt2011chiral}; application of improved and more rigorous many-body methods, including Green's function Monte Carlo \cite{carlson1987green,pudliner1995quantum,pieper2005quantum} 
and coupled cluster\cite{Coester1958421,PhysRevC.59.1440,PhysRevC.69.054320,hagen2010ab} calculations and the no-core shell model \cite{navratil2000large,barrett2013ab,stetcu2013effective}; and improved and more rigorous effective interaction theories such as Okubo-Lee-Suzuki-Okamoto 
theory \cite{okubo1954diagonalization,suzuki1980convergent,suzuki1982unitary,suzuki1983degenerate}, the unitary correlation operator
method \cite{feldmeier1998unitary}, and the similarity renormalization group 
\cite{PhysRevD.48.5863,wegner1994flow,PhysRevC.75.061001,bogner2010low,PhysRevLett.103.082501}.
Powering all these developments has been geometically growing computational power.  Although the problems are far from settled, nuclear 
theorists now routinely perform \textit{ab initio} structure calculations. 

(Herein by \textit{ab initio}  I mean interactions fitted to two- and possibly three-body data, in particular scattering phase shifts and deuteron and 
triton/$^3$He binding energies, and then apply to many-body systems; 
by \textit{phenomenological}  I mean those fitted to many-body excitation spectra and binding energies.)

For the purposes of this paper, I will focus on one particular many-body method, the configuration interaction method, also sometimes called the interacting shell model \cite{BG77,navratil2000large,barrett2013ab,br88,Sh98,ca05}. The basic 
idea is quite simple: in order to solve the many-body Schr\"odinger equation,
\begin{equation}
\hat{H} \Psi(r_1, r_2, \ldots, r_A) = E \Psi(r_1, r_2, \ldots, r_A), \label{eigenproblem}
\end{equation}
one expands in a convenient, orthonormal basis $\{ \Phi_\alpha \}$,
\begin{equation}
\Psi(r_1, r_2, \ldots, r_A) = \sum_{\alpha} c_\alpha \Phi_\alpha(r_1, r_2, \ldots, r_A),
\end{equation}
so the Schr\"odinger equation becomes a matrix eigenvalue equation,
\begin{equation}
\sum_\beta H_{\alpha,\beta} c_\beta = E_\alpha c_\alpha,
\end{equation}
with the Hamiltonian matrix element
\begin{equation}
H_{\alpha, \beta} = \langle \Phi_\alpha | \hat{H} | \Phi_\beta \rangle.
\end{equation}

This deceptively simple beginning masks a multitude of related technical issues. What is meant by a `convenient' basis?  Is it better to have as compact a 
basis as possible to reduce the dimensionality, or to have as simple a basis as possible, to make calculation of  matrix elements efficient? 

For example, 
one often starts with a basis of Slater determinants, or, more accurately, the occupation-representation of Slater determinants, that is, antisymmetrized products of 
single-particle states \cite{BG77,Sh98}. Working with a rotationally invariant Hamiltonian,  we  choose 
 single-particle states that are spherical tensors, that is, have good total angular momentum $j$ and $z$-component $m$;  it is then 
easy to construct Slater determinants with fixed total $z$-component $M$.  This is called an $M$-scheme basis. $M$-scheme bases are  easy to work with, but 
any given $M$-scheme basis state is an admixture of states of different total angular momentum $J$. 

Some configuration-interaction codes instead work in the $J$-scheme, 
with basis states with total angular momentum $J$ fixed rather than $M$.  The resulting dimensionality is typically an order of magnitude smaller. On the surface this 
looks to be a good thing.  But each $J$-scheme basis state itself is a sum of many $M$-scheme states, and so computing the Hamiltonian matrix elements becomes
significantly more time-consuming.  Furthermore, in the $M$-scheme in particular the Hamiltonian matrix is very sparse; depending on the details on the many-body calculation, 
only a few matrix elements out of a million will be nonzero.  Because the eigenvalue problem (\ref{eigenproblem}) for large cases is typically solved using an Arnoldi type 
method such as the Lanczos algorithm \cite{parlett1980symmetric,Lanczos}, based upon matrix-vector multiplications, the computational burden is not just the dimensions of the vectors but the number of 
non-zero matrix elements. $J$-scheme matrices are significantly denser than $M$-scheme ones, and in some cases there are more $J$-scheme nonzero matrix elements 
than in the $M$-scheme case. Furthermore the simplicity of the $M$-scheme allows one to avoid storage of the nonzero matrix elements but to efficiently reconstruct 
the Hamiltonian matrix elements as needed on the fly, primarily by loops over spectators. 

It is not the purpose of this paper to argue the superiority of either the $M$-scheme or the $J$-scheme. Both  have advantages and trade-offs, and over the decades codes have 
been written in both schemes. With modern parallel computing, one can distribute both the wave function vector and the non-zero matrix elements of the many-body 
Hamiltonian (or the arrays used to reconstruct it) over many processors. As far as I can tell, at the time of this writing the largest configuration interaction calculations, 
certainly for nuclear structure physics, had dimensions of roughly 20 billion ($M$-scheme) basis states.   

Such large calculations pose significant computational challenge. Even in single precision, a vector of  $2 \times 10^{10}$ components requires 80 Gbytes of storage. 
Assuming an $M$-scheme sparsity of $10^{-6}$, 
 the nonzero matrix elements would require roughly 5 Petabytes (and indeed, the needle-in-the-haystack problem 
of identifying the nonzero matrix elements efficiently is a non-trivial computational problem), and one matrix-vector multiplication 
would require something like 3 trillion floating-point operations. In the $J$-scheme the vector dimensions are an order of magnitude smaller but the matrices are denser.  

The most crucial question of all is pressed upon us by Richard Hamming's dictum\cite{hamming2012numerical}: ``The purpose of computing is insight, not numbers.''  
What insight, then, do we 
gain from ever-larger matrix eigenvalue problems? The observables we can compare to experiment, after all, are few in number: binding and excitation energies and a handful of 
static and transition matrix elements.

One answer is to look at observables we are unlikely to measure in experiments but yet help us to conceptualize the nuclear wave function. Here we fall back on phenomenology 
from the early days of nuclear physics. The liquid drop model, for example, envisioned  nuclides  as having quadrupole oscillations and deformations 
leading to vibrational and rotational spectra \cite{bohr1998nuclear2}. 
Variants, such as the Goldhaber-Teller model for giant E1 resonances \cite{PhysRev.74.1046}  and the scissors mode for giant M1 resonances 
\cite{PhysRevLett.41.1532,richter1995probing} pictured neutrons and protons oscillating coherently 
against each other.  But how do we connect such classical images to modern many-body theory with discrete numbers of fermions?

The answer is group representation theory.  As described below in section \ref{SU3}, the group SU(3) arises naturally in the description of 
 quadrupole deformations of a nuclear droplet, both static (rotation) and dynamic (vibration).  But one can also construct representations of SU(3) using number-conserving 
 fermion operators\cite{elliott1958collective,harvey1968nuclear}. This means in an 
 appropriate basis with a fixed, finite number of fermions, one can construct irreducible representations of SU(3), and thus in a framework with a fixed number of 
 particles one can arrive rigorously at rotational (and vibrational) band spectra just as in the droplet picture. 

Unlike the rotation group SU(2), SU(3) is  generally not an exact or \textit{dynamical} symmetry of the nuclear Hamiltonian, that is, the generators of $SU(3)$ do not 
commute with the Hamiltonian, in which case eigenstates of the Hamiltonian are eigenstates of the Casimir of SU(3). In fact, SU(3) is strongly mixed, especially by pairing and above all by 
the spin-orbit force \cite{rochford1988survival,PhysRevC.63.014318}. But, as will be discussed in the next section, the mixing is not incoherent but displays strong regularities within spectral ``bands,'' leading to so-called \textit{quasi-dynamical symmetry}  \cite{PhysRevC.58.1539,rowe1999quasi,bahri20003}.  (Another concept is that of \textit{partial} 
dynamical symmetries \cite{PhysRevLett.65.2853,PhysRevLett.84.1866}, but that is beyond our brief here.)

In this paper, I will discuss how one can dissect and illuminate nuclear wave functions using Casimir operators of different groups, that is, operators 
which commute with all the generators of the group and which one can use to separate components.  And not only illuminations, but also construction: one hopes 
to  choose the most important irreps to go beyond the $J$-scheme basis to an even more compact, symmetry-adapted basis for more efficient, most 
physics-based calculations.  

Aside from better, not bigger, calculations, group decompositions of nuclear wave functions are often remarkably robust. The group-theoretical decomposition using decades-old 
phenomenological forces in tiny model spaces often agree surprisingly well with those using the freshest, independently-derived \textit{ab initio} interactions in 
model spaces five or six orders of magnitude larger. I find it heartening that vastly different calculations arrive at similar results: it suggest we are doing the right thing; 
it suggests we really \textit{can} have insight into the structure of nuclear wavefunctions and have a robust picture of them.

\section{A brief history of group theory in nuclear many-body physics, with illustrations}

Applying group representation theory to quantum wavefunctions, including nuclear wave functions, has a long history. This will by no means be an exhaustive review.  
A useful source containing far more technical details is Talmi's book \cite{talmi1993simple}. 

 I  restrict myself to groups applied to the fermion shell model. Group theory is of course at the heart of the 
interacting boson model (IBM) \cite{iachello1987interacting,talmi1993simple}, as well as the fermion dynamical symmetry model (FDSM) arising from SO(8) \cite{wu1986fermion,wu1987fermion,wu1994fermion}, 
primarily designed to connect the fermion shell model to the IBM.  Both the IBM and 
FDSM largely start from group theory to build their models. Because quasi-dynamical symmetries can mimic true dynamical symmetries, and conversely deducing 
symmetries directly from data can mask quasi-dynamical symmetries,
 I prefer to start with 
the fermion shell model with minimal assumptions and then arrive at the group theory. 

The phenomenological interactions I use are in model spaces spanned by a single harmonic oscillator shell: 
Cohen-Kurath (CK) for the $0p$ shell \cite{cohen1965effective}, 
the universal $1s$-$0d$-shell interaction version B (USDB) \cite{PhysRevC.74.034315}, 
and a modified $G$-matrix interaction (GXPF1)  for the $1p$-$0f$ shell  \cite{PhysRevC.65.061301}.  I also use 
an \textit{ab initio} interaction based upon chiral effective field fitted to nucleon-nucleon scattering data 
and deuteron properties \cite{PhysRevC.68.041001}.  
The \textit{ab initio} interaction is softened by the similarity renormalization group to an evolution parameter value of $\lambda = 2.0$ fm$^{-1}$ 
and are expressed in a harmonic oscillator single-particle basis of a given frequency, here  between 16 and 22 MeV depending upon the nuclide. 
The phenomenological interactions do not have a specified radial basis for single-particle states, though it is common place to assume a harmonic oscillator basis; for the calculations  
here the frequency does not matter. All \textit{ab initio} calculations were computed in the no-core shell model (NCSM) formalism \cite{navratil2000large,barrett2013ab,stetcu2013effective}, with the model space defined by 
the harmonic oscillator frequency as well as $N_\mathrm{max}$ which denotes the maximum excitation, in harmonic oscillator quanta, allowed above the lowest 
shell model configuration. 

All the many-body calculations described here used the {\tt BIGSTICK} configuration-interaction code \cite{BIGSTICK}. 
The method of decomposition is described below in Section \ref{lanczos}.

\subsection{Spin-orbit}

\begin{table}
\caption{Analysis of ground state wave functions for ``simple'' nuclides in the $0p$, $1s0d$, and $1p0f$ spaces.}
{\begin{tabular}{|c|c|r|r |}\hline
Nuclide & space & \multicolumn{2}{c|}{g.s.} \\
\hline
$^8$He & $0p$ &  $53\% (0p_{3/2})^4$ & $96\% L=0$ \\
$^{12}$C & $0p$ &  $37\% (0p_{3/2})^8$ & $82\% L=0$ \\
\hline
$^{22}$O & $1s0d$ &  $75\% \,\,(0d_{5/2})^6$ & $38\% L=0$ \\
$^{24}$O & $1s0d$ &  $91\% \,\,(0d_{5/2})^6(1s_{1/2})^2$ & $34\% L=0$ \\
$^{28}$Si & $1s0d$ &  $21\% \,\,(0d_{5/2})^12$ & $36\% L=0$ \\
$^{32}$S & $1s0d$ &  $29\% \,\,(0d_{5/2})^12(1s_{1/2})^4$ & $34\% L=0$ \\
\hline
$^{48}$Ca & $1f0p$ &  $90\% \,\,(0f_{7/2})^8$ & $20\% L=0$ \\
\hline
\end{tabular}
}
\label{LSvsjj}
\end{table}

The nuclear Hamiltonian is invariant under rotation, but ever since the Dirac equation it's been clear that angular momentum has two pieces, orbital angular momentum 
$\vec{l} = \vec{r} \times \vec{p}$ and intrinsic spin $s$, such that, for a given particle, the angular momentum $j$ is a tensor sum of $l$ and $s$. 
 In the electronic structure of atoms the spin component makes only a tiny contribution through magnetic interactions 
to the Hamiltonian, and so the orbital angular momentum and spin nearly decouple. This naturally leads to \textit{L-S} or \textit{Russell-Saunders} coupling scheme, where one 
carries out a tensor sum of all the $l$s for the particles to get the total orbital angular momentum $L$, and separately sum the individual spins to get the total spin $S$; then 
$L$ and $S$ couple to form the total angular momentum $J$.  

In medium- to heavy-nuclei,  large spin-orbit splitting, which arises naturally in relativity, energetically separates orbits with the same $l$ but different $j$, 
leading to the $j$-$j$ coupling scheme: for each particle one couples $l$ and $s$ to $j$ and then sum directly to total $J$.  In fact it was only 
when Maria Goeppert-Mayer \cite{mayer1948closed,PhysRev.75.1969,PhysRev.78.16} and others \cite{PhysRev.75.1766.2,PhysRev.75.1968.2} realized that strong spin-orbit 
splitting of levels could explain the closure at ``magic numbers'' as well as the magnetic dipole moments of doubly-magic $\pm 1$ nuclides that $j$-$j$ 
coupling took hold in nuclear physics. 

For open $p$-shell nuclides, it was not immediately clear which coupling was superior: $L$-$S$ \cite{PhysRev.51.95,PhysRev.51.597} 
or $j$-$j$ \cite{flowers1952studies,PhysRev.88.804}. To give a preview of the answer, let's consider some phenomenological calculations.  Table \ref{LSvsjj} show several semi-magic nuclides. In the simplest $jj$ coupling picture, the ground states should be described 
by filled orbits. Indeed, for the neutron-rich semi-magic nuclei in the $1p0f$ and $1s0d$ spaces we see this is the case. For the $0p$ nuclei, however, they are poorly 
described by filling the $0p_{3/2}$ orbit, but better described by a $L=0$ state.

\begin{figure}
\centerline{\includegraphics[width=8.8cm]{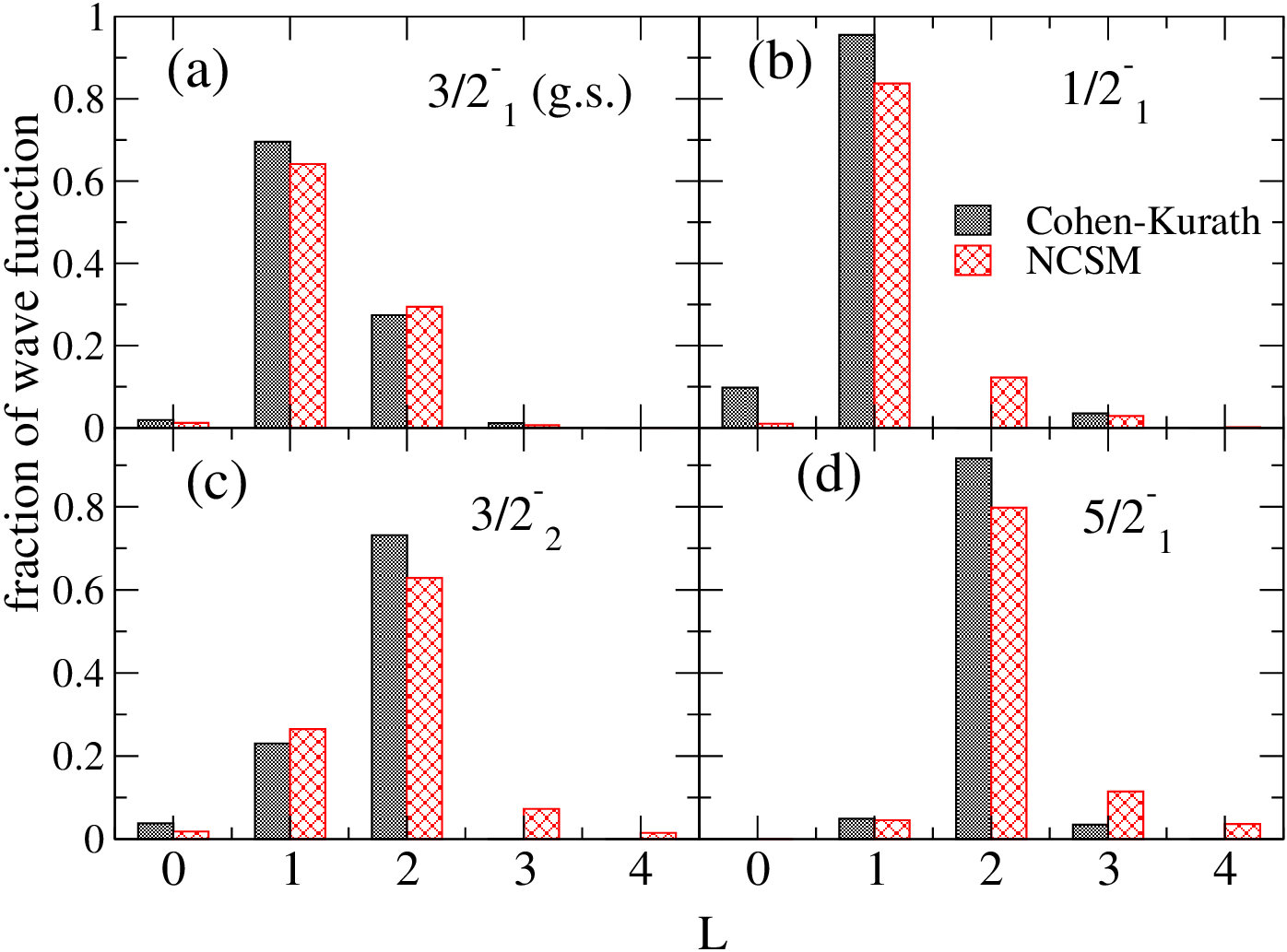}}
\caption{$L$-decomposition of the $0p$-shell nuclide $^{11}$B}
\label{B11_Ldecomp}
\end{figure}

To illustrate, let's start in the $0p$ shell with a nontrivial nuclide, $^{11}$B, using both the phenomenological Cohen-Kurath interaction (dating from 1965!) 
in the $0p$ shell assuming an inert $^4$He core, which has a basis dimension of 62 $M$-scheme states,
 and a no core shell model (NCSM) with a modern \textit{ab initio} interaction, computed in a harmonic oscillator basis with oscillator frequency $\hbar\Omega=22$ MeV 
 and allowing excitations up to $N_\mathrm{max}=6$. .
 
 Figure \ref{B11_Ldecomp} shows the decompostion of the first four states into $L$-components, 
There are a number of things to note here. First, 
the very good, even surprising agreement between the phenomenological calculation with the Cohen-Kurath interaction and the \textit{ab initio} NCSM calculation. 
Second while the $1/2^-$ and $5/2^-$ states are dominated by a single $L$-value, both the first and second $3/2^-1$ states have nontrivial--and contrasting--secondary components.
Not shown  is the $S$-decomposition, which is nearly trivial with nearly all the wavefunction in the $S=3/2$ irrep.  This is not surprising, given that the spin-orbit splitting is relatively 
weak in the $0p$ shell.

\begin{figure}
\centerline{\includegraphics[width=8.8cm]{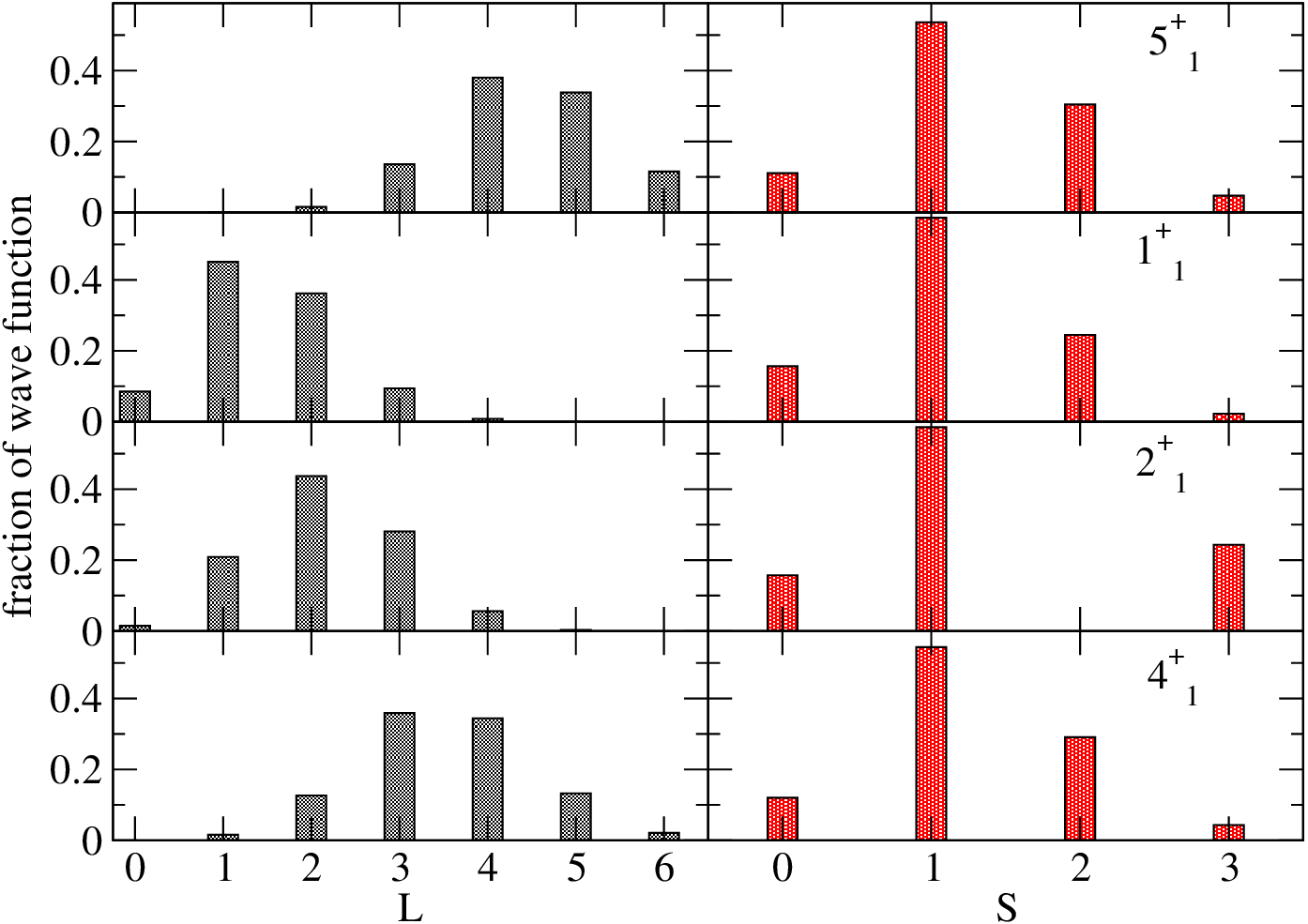}}
\caption{$L$- (lefthand side) and $S$-decomposition (righthand side) of the $1p$-$0f$-shell nuclide $^{48}$V,
for the first four states.}
\label{V48_LSdecomp}
\end{figure}

Therefore for contrast Fig.~\ref{V48_LSdecomp} shows the $L$- and $S$-decomposition for the first four states of the odd-odd nuclide $^{48}$V in the $1p0f$ shell. 
Because an NCSM calculation is not currently practical for this nuclide, I only show results from a phenomenological calculation. 
While relatively more complicated than for $^{11}$B, the wavefunction nonetheless show regularity and simplicity: the $S$-decompositions, while fragmented, are strikingly similar, 
especially for the $4^+$ (g.s.), the $1^+$, and the $5^+$ states.  The $L$-decompositions also exhibit similarity across states, 
differing mostly coherent shifts along the $L$-axis, suggesting 
an intrinsic shape being spun up. We'll see more of this behavior when we look at rotational bands in more depth below. 

An $L$-$S$ description of nuclei never really took direct hold because of the strong spin-orbit splitting, although recent work\cite{PhysRevC.91.034313} showed it nonetheless
illuminates $0p$-shell wavefunctions. But an SU(3) basis for nuclei implicitly includes $L$-$S$ decompositions because SU(3) is cast in the spatial part of the wave function and 
separated from the spin part. 

One can of course go further, and decompose for the proton portion of $L$ or the neutron portion of $S$, but to the sake of a streamlined narrative I won't show such 
detailed decompositions.

\subsection{Isospin and SU(4) }

Heisenberg's introduction of the concept of isospin \cite{heisenberg1932structure} was one of the first applications of group theory to nuclear physics. In this case 
isospin was a straightforward analog to spin.   Isospin is not  exactly conserved but is  broken at only the few percent level and so can be treated as an 
exact symmetry. The most immediate evidence for  isospin symmetry in nuclear spectra is through mirror symmetry (i.e., the same level scheme for nuclei for 
the same $A$ and opposite $T_z = (N-Z)/2$, e.g., $^{13}$C and $^{13}$N), and isobar analog states (levels with $T > |T_z|$ mirroring those in nuclei with the same $T$ but
with $T=T_z$, i.e., $^{12}$C has low-lying levels with $T=0$ but also has $T=1$ levels mirroring those in $^{12}$B and $^{12}$N).

\begin{figure}
\centerline{\includegraphics[width=8.8cm]{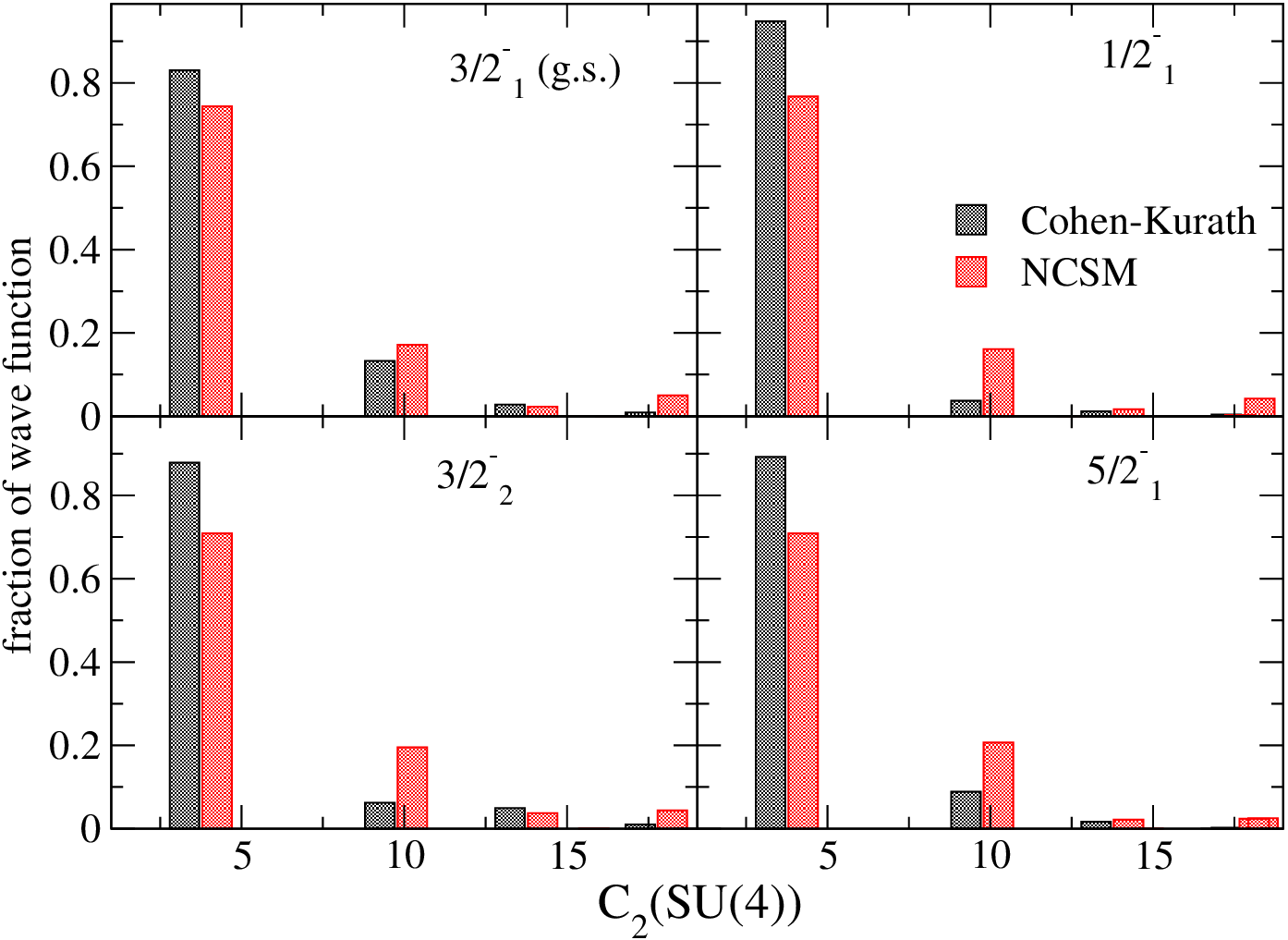}}
\caption{$SU(4)$-decomposition of the $0p$-shell nuclide $^{11}$B}
\label{B11_SU4}
\end{figure}

Because one can decompose the group $SU(4) \supset SU(2) \times SU(2)$, Wigner suggested \cite{PhysRev.51.106,hecht1969wigner}  looking for an $SU(4)$ symmetry built upon 
$SU_S(2) \times SU_T(2)$, sometimes called a \textit{supermultiplet}.  
  The irreps of SU(4) are labeled by the quantum numbers $P,P^\prime$, and $P^{\prime \prime}$, which arise from the Young tableaux \cite{talmi1993simple},
  which is found by the Casimir operator
 \begin{equation}
 C_2(SU(4)) = \vec{S}^2 + \vec{T}^2 + 4 \vec{S}^2 \vec{T}^2,
 \end{equation}
 which has eigenvalues
 \begin{equation}
 P(P+4) +P^\prime(P^\prime+2) +\left( P^{\prime \prime}\right)^2
 \end{equation}
 In the highest weight states, $P=S$ and $P^\prime=T$. 
 
\begin{figure}
\centerline{\includegraphics[width=8.8cm]{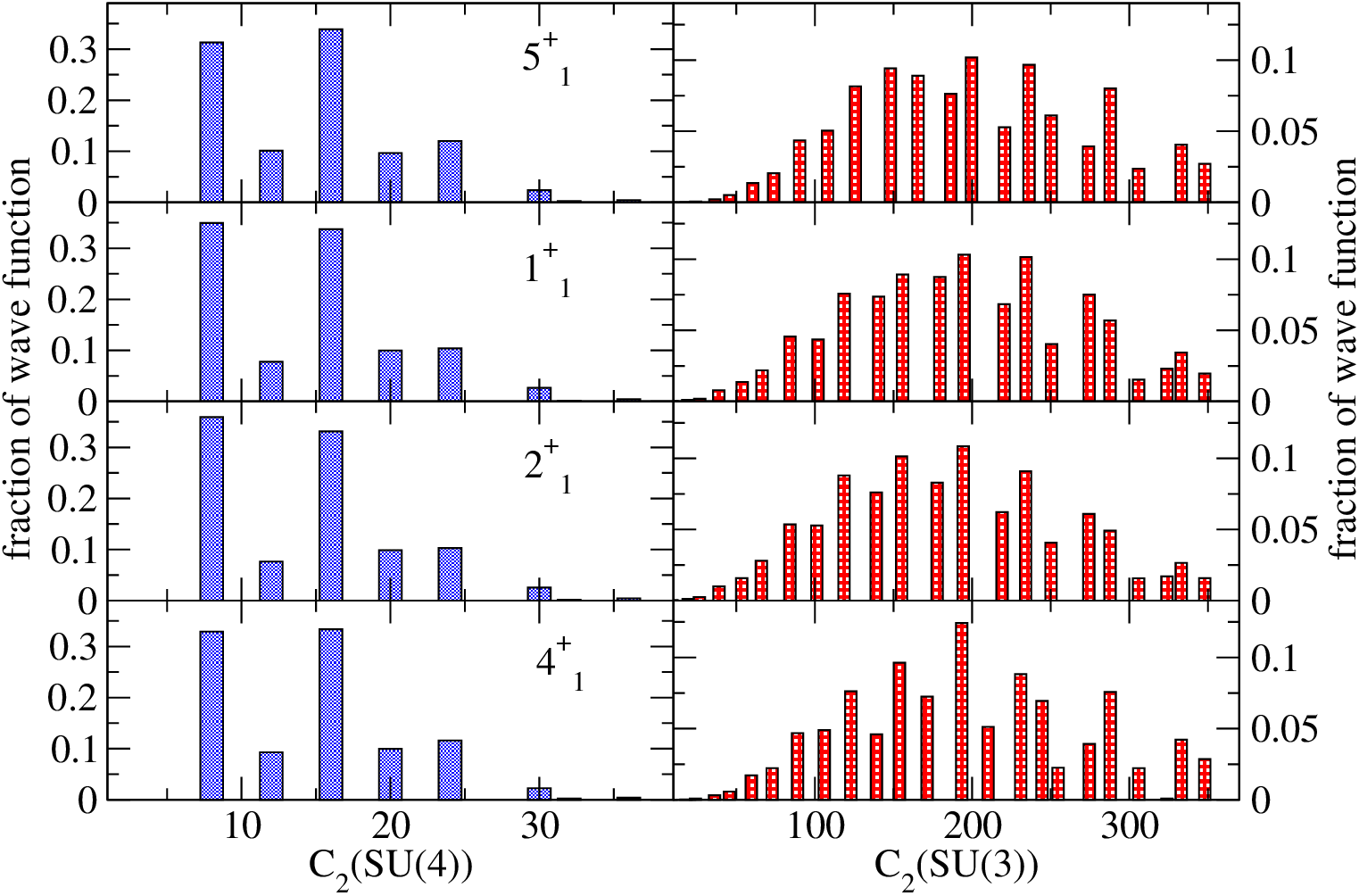}}
\caption{$SU(4)$ (left) and $SU(3)$ (right) decomposition of the first four states of the $1p0f$-shell nuclide $^{48}$V}
\label{V48_SU4SU3}
\end{figure}

 Wigner's suggestion led to  efforts to look for applications where SU(4) symmetry is 
 approximately valid \cite{franzini1963validity,hecht1974spectral}.
Of course, from the very start this search looks doomed, as the nuclear force in the spin-triplet, isospin-singlet channel, 
which gives us the bound deuteron, is very different from the spin-singlet, isospin-triplet channel.
Indeed at low energies, spin-isospin composition of medium- to heavy-nuclei show significant fragmentation of the wave function over many irreps 
 \cite{PhysRevC.47.623,Frazier1997}.   We see fragmentation in Fig.~\ref{B11_SU4} for 
 $^{11}$B (again with Cohen-Kurath and NCSM calculations) and for $^{48}$V in the left-hand panels of Fig.~\ref{V48_SU4SU3}; both cases also 
 exhibit evidence of coherent quasi-dynamical symmetry.

\subsection{Pairing, seniority, and quasi-spin}

Racah introduced the seniority scheme  \cite{racah1942theory,racah1952farkas,flowers1952studies} to aid in the understanding of  complex 
spectra, where the seniority quantum number $\nu$ denotes the number of `unpaired' nucleons, and where the `pairing force' came to be understood as 
an approximation for a zero-range $\delta$-force \cite{racah1952pairing}.  The introduction of the quasi-spin scheme \cite{kerman1961pairing} made clear one 
could represent seniority through SU(2).  

The problem with Racah's seniority is that the pairing operator is equally weighted over all orbits, a scheme immediately and badly broken by single-particle 
energies.  One can then appeal to generalized seniority as a description of states \cite{talmi1993simple}, but one loses the quasi-spin formalism and 
the power of group theory.

\subsection{SU(3) and beyond}

\label{SU3}

Unlike atoms, atomic nuclei deform quite easily, and the liquid drop model, 
provides a good phenomenological description of rotational and vibrational bands. If one thinks about a quadrupole-deformed body whose surface is 
given by 
\begin{equation}
R(\theta, \phi) = R_0 \left( 1 + \sum_m \alpha_{2,m} Y_{2,m}(\theta,\phi) \right),
\end{equation}
where the $Y_2m$ are the spherical harmonics for $l=2$. In the Bohr-Mottelson model \cite{bohr1998nuclear2}, the dynamical amplitudes $\alpha_{2,m}$ are 
quantized. Of the five degrees of freedom, two are the traditional deformation parameters $\beta, \gamma$ which are roughly deformation magnitude and triaxiality, respectively, 
while the remaining three are orientation through the Euler angles.  

It turns out the five quadrupole operators $Q_m = r^2 Y_{2m}(\theta,\phi)$ and the three generators of orbital angular momentum $\vec{L} = \vec{r} \times \vec{p}$ 
have closed commutation relations and form the eight generators of the group SU(3). The sheer number of papers on applications of SU(3) preclude a 
complete review, but the basic story is straightforward. 
The key step was Elliott's introduction \cite{elliott1958collective,harvey1968nuclear} of a finite representation of SU(3) using one-body 
fermion operators for $\vec{Q}, \vec{L}$. This allowed a direct connection between the well-known phenomenology of deformation and 
the fermion shell model.

\begin{figure}
\centerline{\includegraphics[width=8.8cm]{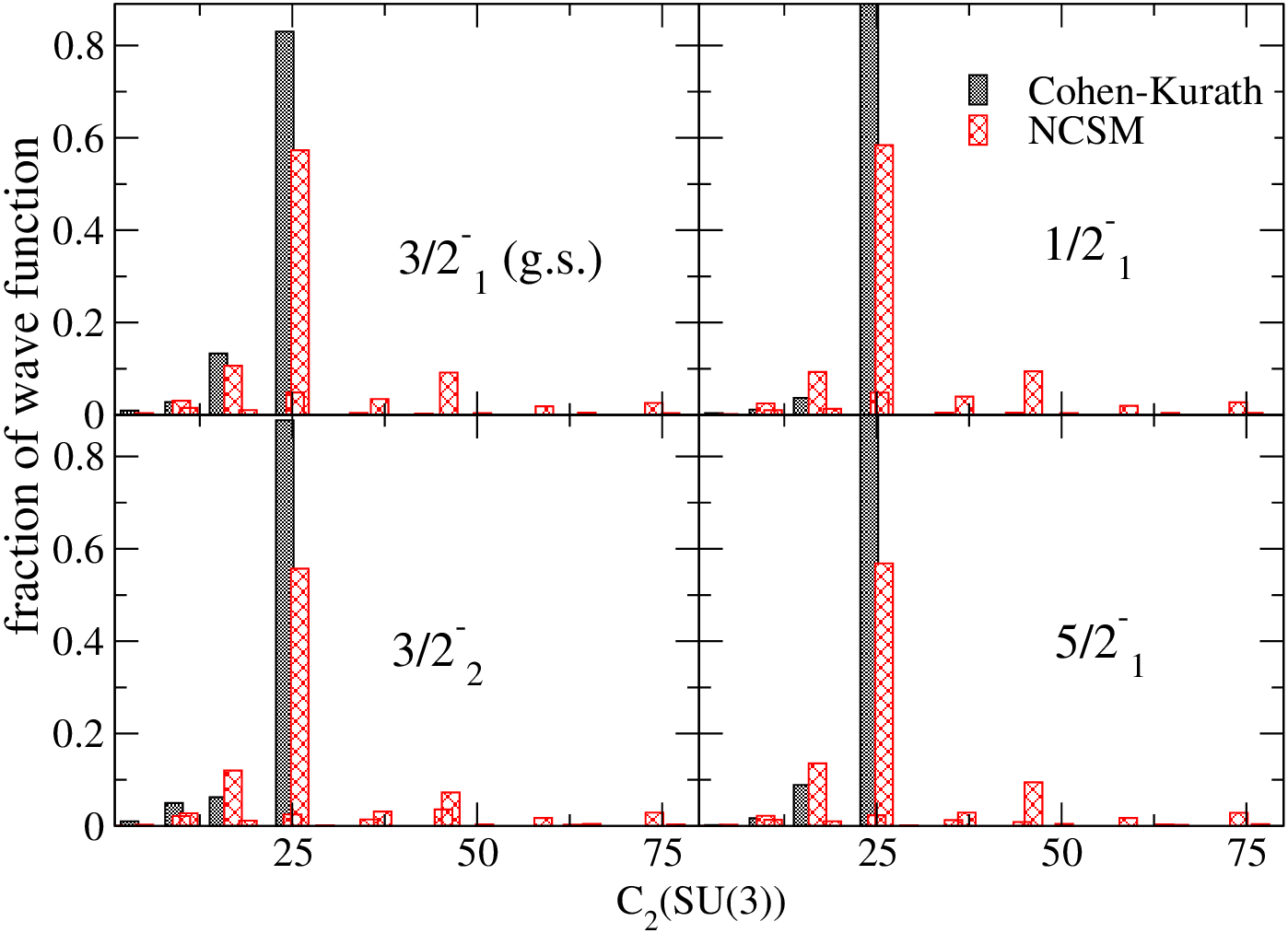}}
\caption{$SU(3)$-decomposition of the $0p$-shell nuclide $^{11}$B}
\label{B11_SU3}
\end{figure}

The irreps of SU(3) are labeled by the quantum numbers  $\lambda$ and $\mu$ via their Young tableaux \cite{talmi1993simple}, and 
which can be interpreted in terms of the standard deformation parameters $\beta$ and $\gamma$ (see Figure 1 in Ref.~\cite{Bahri1995171}).
The second-order Casimir operator, 
\begin{equation}
C_2(SU(3)) = \frac{1}{4} \left( \vec{Q} \cdot \vec{Q} + 3 L^2 \right),
\end{equation}
where
\begin{equation}
Q_m= \sqrt{\frac{4\pi}{5}} \left ( \frac{r^2}{b^2} + b^2 p^2 \right ) Y_{2m}(\theta, \phi)
\end{equation}
is the (dimensionless) so-called Elliott quadrupole operator, whose matrix elements are nonzero only within a major harmonic oscillator shell ) 
has eigenvalues $\lambda^2 + \lambda \mu + \mu^2 + 3\lambda + 3\mu$.  
One can distinguish between different combinations of $\lambda$ and $\mu$ by including the third-order Casimir, although for simplicity here I 
only use the second-order Casimir.

For realistic interactions, SU(3) is not a dynamical symmetry because the generators of SU(3) do not commute with the Hamiltonian. 
Two important sources of SU(3) symmetry breaking is spin-orbit splitting \cite{rochford1988survival,PhysRevC.63.014318} and pairing \cite{Bahri1995171}. 
In the limit of very strong spin-orbit splitting one can take the so-called pseudo-SU(3) limit \cite{Arima1969517,raju1973search}
in phenomenological calculations. 

Decomposition of $^{11}$B into SU(3) irreps is shown in Fig.~\ref{B11_SU3}. Because of the weak spin-orbit splitting, SU(3) is nearly a good symmetry here and all 
these states are dominated by the same irrep. The \textit{ab initio} NCSM calculation is more fragmented than the phenomenological Cohen-Kurath results, 
but that is largely because it includes higher harmonic oscillator shells.  By contrast, the SU(3) decomposition of the $1p0f$ nuclide $^{48}$V, 
as shown on the right-hand side of Fig.~\ref{V48_SU4SU3} is highly fragmented, 
in large part due to the large spin-orbit force, although remnants of quasi-dynamical symmetry remain.  Rotational $1p0f$ spectra, such as for $^{44}$Ti and $^{48}$Cr, 
are still highly fragmented but nonetheless 
exhibit the coherent structures \cite{PhysRevC.63.014318} of quasi-dynamical symmetry \cite{PhysRevC.58.1539,rowe1999quasi,bahri20003}.

While the Elliott SU(3) model provides a direct connection between deformation and the fermion shell model, it is limited by the fact the operators do not connect 
across major harmonic oscillator shells. 
A natural generalization of SU(3) is the symplectic group or Sp(3,R) \cite{PhysRevLett.38.10,rosensteel1979algebraic,park1984shell}, which adds raising and lowering operators to connect 
different oscillator shells. Furthermore, one can naturally separate out spurious center-of-mass motion in the symplectic model, meaning one can have a translationally 
invariant theory. 

Despite these appealing features, work using the symplectic representation has been limited, in no small part due to the difficulty of 
the group theory. Nonetheless there have been several promising developments and initial applications in this direction, as described below in section \ref{Symp}.

\section{Lanczos decomposition of wave functions}

\label{lanczos}

Before giving more examples, let me explain a useful technique for carrying out these decompositions.

Given eigenstates $| \Psi \rangle$  of the Schr\"odinger equation (\ref{eigenproblem}), we want to decompose them according to 
irreducible representations of a group. To be specific, we divide up any space into subspaces labeled by the eigenvalues $\gamma$ of a Casimir operator $\hat{\cal C}$ of a 
group, that is, we label states by
\begin{equation}
\hat{\cal C} | \Gamma \rangle  = \gamma | \Gamma \rangle.
\end{equation}
While we generally have a physical or mathematical motivation for looking at group irreps, there is also a very practical reason: in general a Casimir has a highly 
degenerate spectrum, by which I mean, relative to the dimension of any model space, there are only a small number of distinct eigenvalues $\gamma$, with 
with a large number of states associated with each eigenvalue. Dividing a space into subspaces labeled by $\gamma$ is thus a straightforward task. 

Because each irrep has many states, in principle we need some additional index $i$ to label the states $| \Gamma; i \rangle$ of a given irrep. How to best do this can be a 
nontrivial task. It will turn out that problem won't concern us. 

Instead, for a given eigenstate $ | \Psi \rangle$ of the Hamiltonian, one can define the fraction in a given irrep by
\begin{equation}
\mathrm{frac}(\gamma) = \sum_{i \in \gamma} |\langle \Gamma ; i | \Psi \rangle |^2 .
\end{equation}
At first this looks very daunting: the number of states in any given irrep can be very large.  Fortunately for us, the Lanczos algorithm \cite{Lanczos} comes to the rescue. 

For those readers not fluent with the Lanczos algorithm, used in its standard form to find extremal eigensolutions, let me describe it.  
Starting from an initial vector $| v_1 \rangle,$ often called the \textit{pivot} and using a Hermitian (typically real, symmetric) matrix $\hat{H}$, the Lanczos algorithm carries out a 
sequence of matrix-vector multiplications to iteratively constructs an orthonormal set of basis vectors:
\begin{equation}
\begin{array}{lclllll}
\hat{H}| v_1 \rangle & = & \alpha_1 | v_1\rangle  + &  \beta_1 | v_2 \rangle & & & \nonumber \\
\hat{H}| v_2 \rangle & = & \beta_1 | v_1\rangle +  &  \alpha_2 | v_2 \rangle + &  \beta_2 | v_3 \rangle  & & \\
\hat{H}| v_3 \rangle & = &  & \beta_2 | v_2\rangle  + &  \alpha_3 | v_3 \rangle + & \beta_3 | v_4 \rangle &   \\
\hat{H}| v_4 \rangle & = & & &  \beta_3 | v_3\rangle  + & \alpha_4 | v_4 \rangle + &  \beta_4 | v_5 \rangle  \\
\ldots                       &    &  & &                                       &                                       &
\end{array}
\end{equation}
The next step is to find the eigenpairs of the resulting tridiagonal matrix. The downside of the Lanczos algorithm is that the orthogonality of the Lanczos 
vectors $\{ |v_i \rangle \}$ must be numerically enforced; hence when finding all the eigenpairs of a matrix, one typically uses the Householder algorithm, 
which also constructs an intermediate tridiagonal matrix through a more stable, though less straightforward, sequence of unitary transformations. 

The Lanczos algorithm has its uses, however.  Suppose one truncates the tridiagonal matrix at the $n$th iteration, with $n \ll N $, the dimension of the full space.
By the variational principle one can easily see the eigenvalues of  the truncated tridiagonal are bounded by the eigenvalues of $\hat{H}$ in the full space. 
Furthermore, the extremal eigenvalues of the truncated tridiagonal quickly converge, as a function of $n$, to the extremal eigenvalues of $\hat{H}$.
This is very useful in applications such as nuclear structure physics where typically one doesn't need all the eigenpairs but is concerned only with 
the lowest few states. While the exact convergence depends on the system, the ground state typically converges in 50-70 iterations and the lowest ten states 
in less than 300 iterations. The Householder algorithm goes like $N^3$ while Lanczos goes like $n N^2$. Furthermore, as the Lanczos algorithm and its 
variants are built 
around matrix-vector multiplication followed by orthogonalization, it is easy to conceptualize. 
 The Lanczos algorithm is simple, beautiful, and powerful, with an enormous literature devoted to it.

The Lanczos algorithm has applications beyond simply finding eigenpairs. To understand this, let's look in depth at how one constructs the eigen\textit{vectors}. 
We want to find 
$$
\hat{H} | \Psi_a \rangle = E_a | \Psi_a \rangle.
$$
In the Lanczos algorithm, we first construct the tridiagonal matrix $\mathbf{L}^{(n)}$ which is truncated to dimension $n$. It's important to note that 
$\mathbf{L}$ is in the space of the Lanczos vectors $\{ | v_i \rangle \}$, also called the \textit{Krylov subspace}, that is
\begin{equation}
L_{ij} = \langle v_i | \hat{H} | v_j \rangle.
\end{equation}
$\mathbf{L}^{(n)}$ in turn is diagonalized
\begin{equation}
\mathbf{L}^{(n)} | \lambda_a \rangle = \tilde{E}_a |\lambda_a \rangle
\end{equation}
where the extremal $\tilde{E}_a \rightarrow E_a$ as $n$ increases. Because $\mathbf{L}$ is in the Krylov subspace, in the computer the eigenvectors 
$| \lambda_a \rangle = \sum_i d_{i,a} |v_i \rangle$.  By knowing the Lanczos vectors in the original basis, one can then reconstruct $|\Psi_a \rangle$ in 
the original basis. 

But suppose the pivot, $|v_1 \rangle$, is a special vector, such that we want to know the overlap $\langle v_1 | \Psi_a \rangle$. This one can just read off 
directly as $d_{1,a}$, that is, $\mathrm{frac}(\gamma) = | d_{1,g}|^2$ where $g$ labels the eigenvector with eigenvalue $\gamma$.  With a judicious choice of 
Lanczos iterations, roughly the number of unique eigenvalues for the Casimir, each eigenvalue is associated with only one eigenvector.

This technique was originally introduced to calculate strength functions
\cite{caurier1990full,PhysRevLett.74.1517,PhysRevC.59.2033,PhysRevC.72.065501} but was later adapted for symmetry-guided decompositions 
\cite{PhysRevC.63.014318,PhysRevC.91.034313}.

\section{Rotational bands}

\begin{figure}
\centerline{\includegraphics[width=8.8cm]{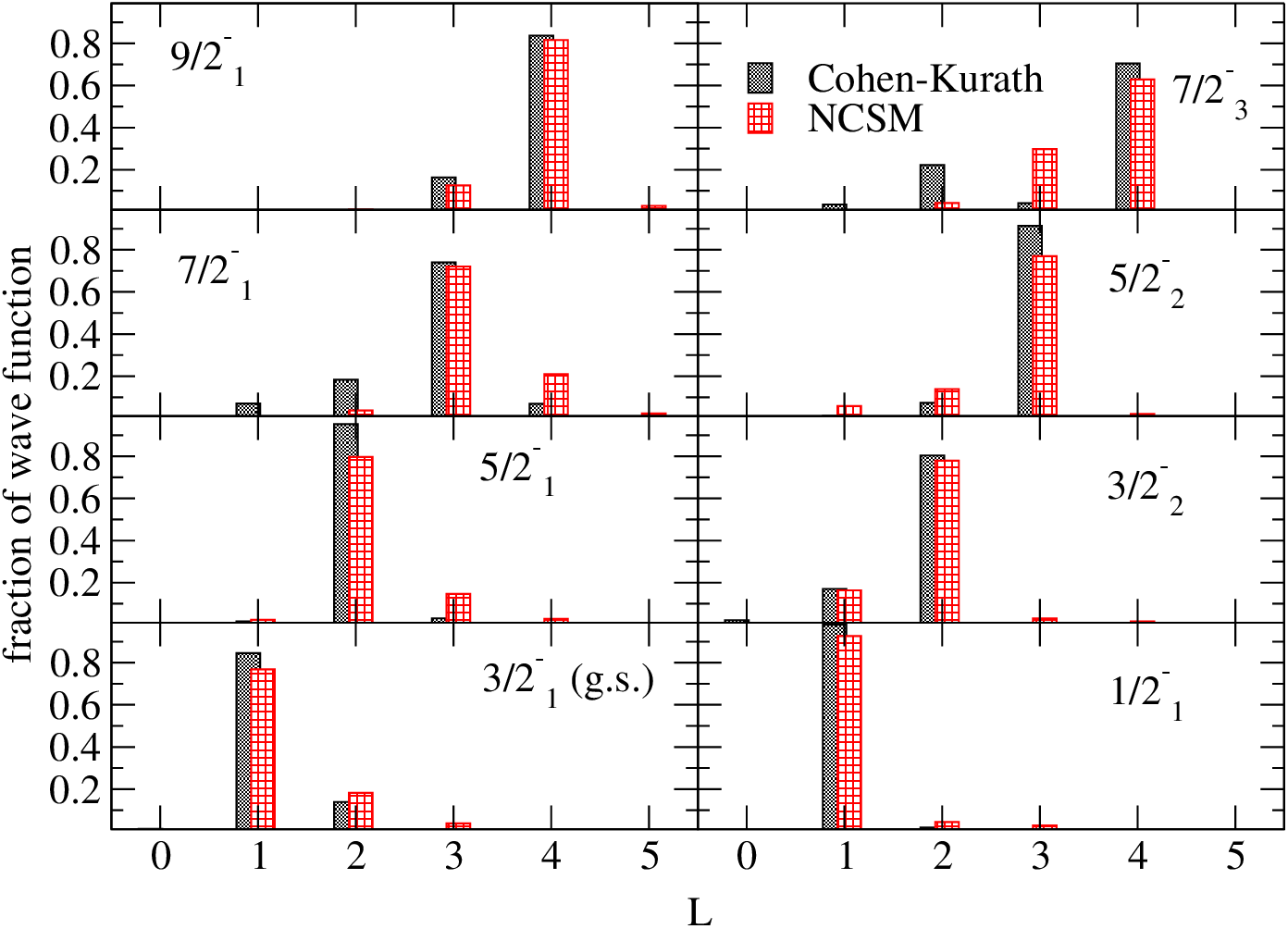}}
\caption{$L$ decomposition of the ground state (left side) and first excited  (right side) rotational bands of $^{9}$Be, using both phenomenological Cohen-Kurath interaction 
(black, shaded) and \textit{ab initio} NCSM spaces and interactions. }
\label{Be9_Ldecomp}
\end{figure}

Historically, one of the strongest motivations for group-theoretical approaches to nuclear structure has been regular band structures, not only the pattern of 
excitation energies but also electromagnetic transitions.  These patterns led first to the liquid drop picture and its quantized incarnation the 
Bohr-Mottelson model and later to the Nilsson model\cite{bohr1998nuclear2}, the SU(3) model\cite{elliott1958collective,harvey1968nuclear}, and the interacting Boson Model and other algebraic models\cite{iachello1987interacting,talmi1993simple}.  As described above, detailed calculations showed that we seldom have pure dynamical symmetries, 
but instead quasi-dynamical symmetries are commonplace. 
Of all the band structures, rotational bands built upon static deformations are the most common, and so I would like to showcase further two more examples, 
$^9$Be and $^{20}$Ne.

\begin{figure}
\centerline{\includegraphics[width=8.8cm]{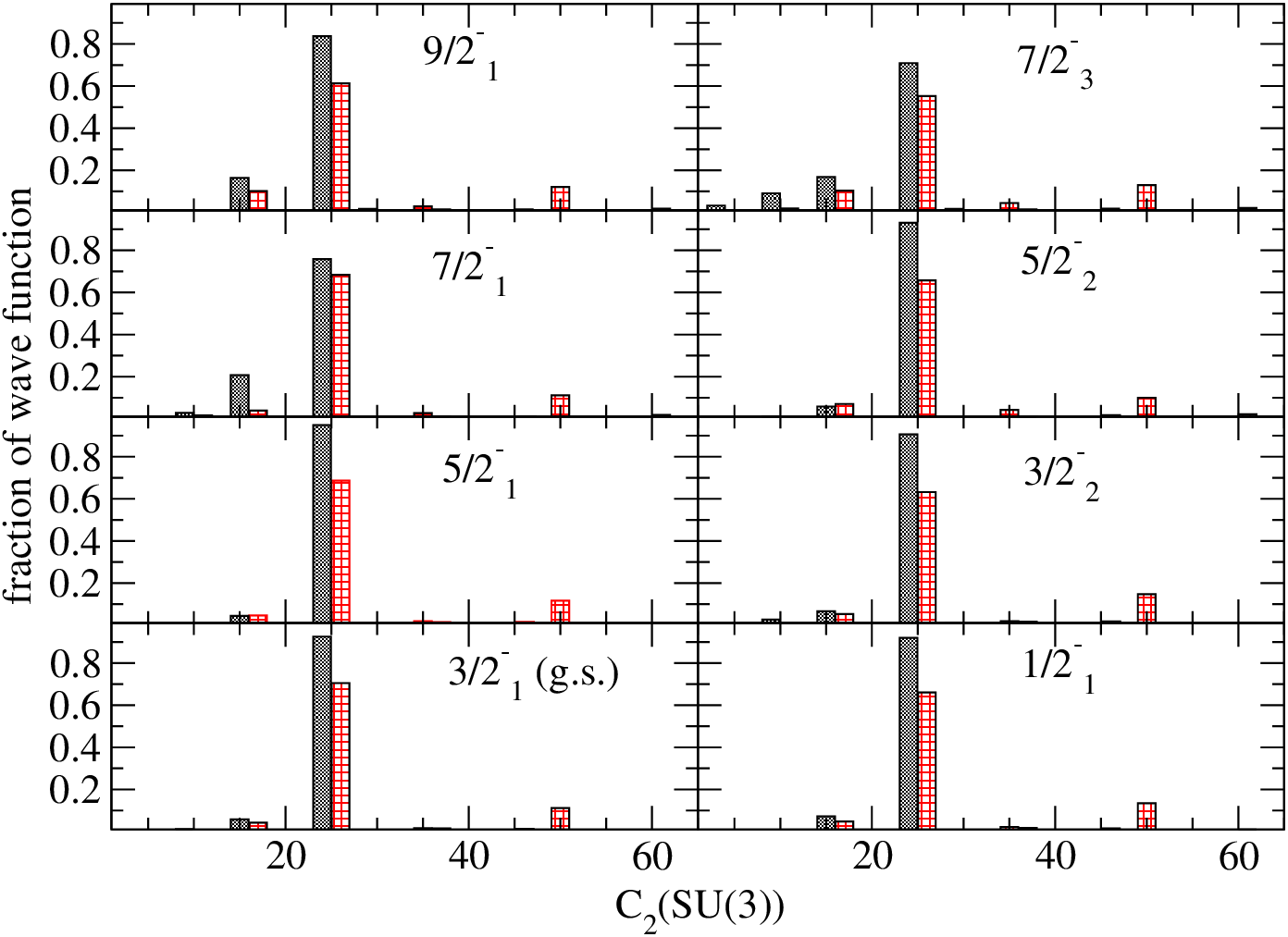}}
\caption{$SU(3)$ decomposition of the ground state (left side) and first excited  (right side) rotational bands of $^{9}$Be, using both phenomenological Cohen-Kurath interaction 
(black, shaded) and \textit{ab initio} NCSM spaces and interactions. }
\label{Be9_SU3}
\end{figure}

 $^9$Be I  compute both in the $0p$ space with the phenomenological Cohen-Kurath interaction, with dimension of 62 $M$-scheme states (this happens to be the 
 same as $^{11}$B as they are, in this space, particle-hole conjugates of each other), and in the NCSM with $N_\mathrm{max}=6$ 
excitations allowed in a harmonic oscillator basis with frequency $\hbar \Omega = 20$ MeV, with a dimension of 20 million states; rotational bands in this nuclides and its sister isotopes has been studied in depth in 
\textit{ab initio} calculations \cite{PhysRevC.91.034313,caprio2013emergence,PhysRevC.91.014310}. I show both the ground state and excited state bands. 
Fig.~\ref{Be9_Ldecomp} shows the wavefunctions are dominated by a single $L$ value increasing steadily. Both bands are nearly exclusively $S=1/2$.
The SU(3) decomposition, shown in Fig.~\ref{Be9_SU3}, is similarly dominated by a single, consistent irrep for both bands.  The natural interpretation is a fixed intrinsic 
shape being spun up, with different couplings between the $S=1/2$ component and the $L$ component.  The dominance by a single $L$ value and a single SU(3) irrep 
can be understood as the spin-orbit force is still too weak to dramatically fragment the distribution. 

This decomposition can even be used to identify members of a band.  The ground state band claims the first $7/2^-$ state as a member, while the excited state band 
takes not the second but the third $7/2^-$ state. The second $7/2^-$ has a quite different structure, dominated by $S=3/2$ (not shown).  This differentiation was also found 
elsewhere by looking at systematics of electric quadrupole and magnetic dipole moments and transitions  \cite{caprio2013emergence,PhysRevC.91.014310}.

\begin{figure}
\centerline{\includegraphics[width=8.8cm]{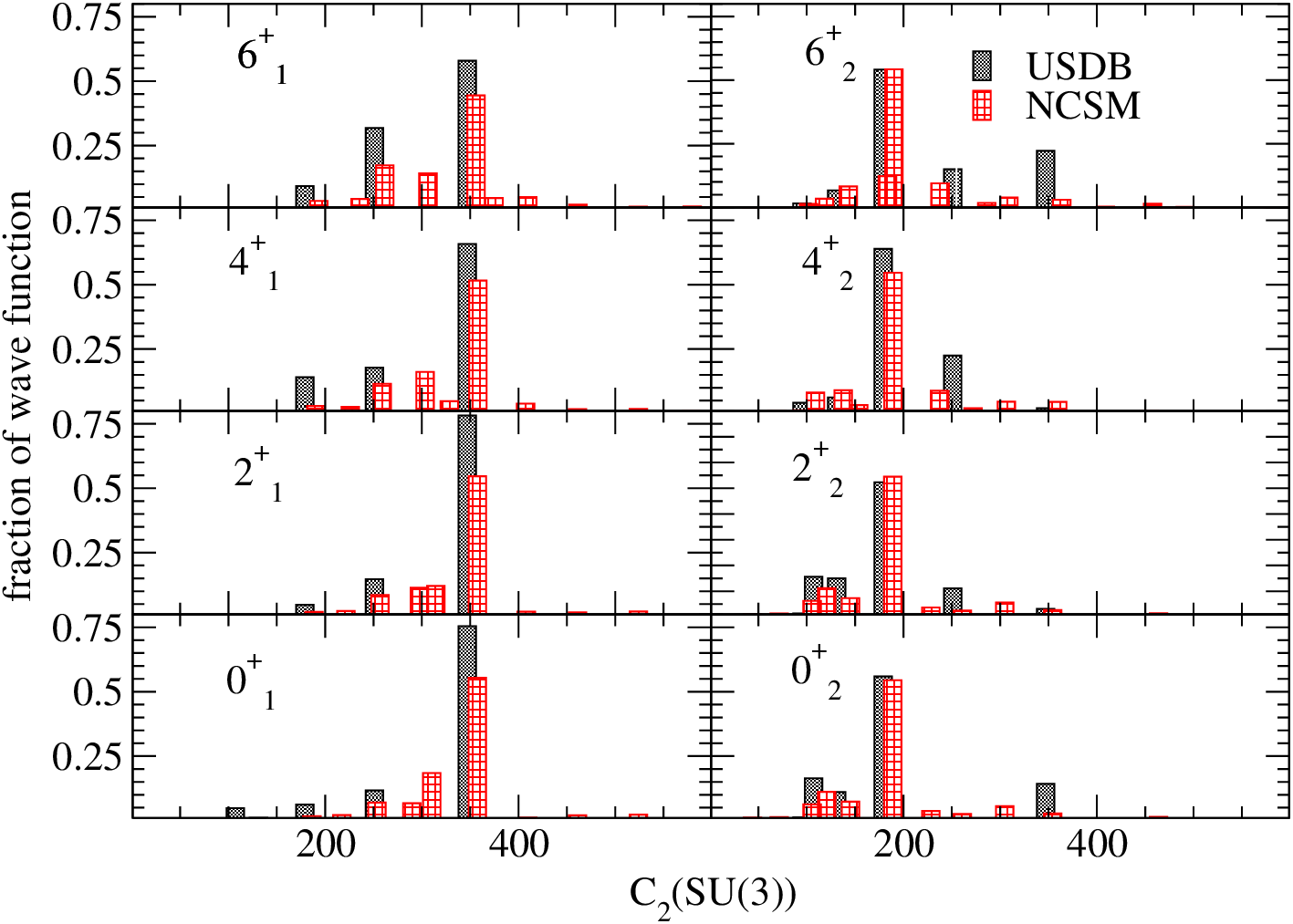}}
\caption{$SU(3)$ decomposition of the ground state (left side) and first excited  (right side) rotational bands of $^{20}$Ne, using both phenomenological USDB interaction 
(black, shaded) and \textit{ab initio} NCSM spaces and interactions. }
\label{Ne20SU3}
\end{figure}

$^{20}$Ne I compute in the $1s0d$ space with the phenomenological 
USDB interaction \cite{PhysRevC.74.034315} with an $M$-scheme dimension of 640 states, 
and in the NCSM with $N_\mathrm{max} = 4$ excitations allowed in a harmonic oscillator basis with frequency $\hbar\Omega=16$ MeV, dimension of  75 million basis states. 
Fig.~\ref{Ne20SU3} shows its decomposition into SU(3) irreps. While like $^9$Be it is also dominated by a single irrep, unlike $^9$Be the ground state and excited bands 
are dominated by very different intrinsic shapes. 

In both cases both the phenomenological and the \textit{ab initio} calculations give similar results, despite vast differences in origin and in model space size. Indeed, 
for the most part such decompositions are robust against choice of model space size, basis parameters (i.e., harmonic oscillator frequency $\hbar\Omega$), and 
even SRG evolution \cite{PhysRevC.91.034313}.  It also seems likely we have just begun to apply these tools to understanding nuclear wavefunctions.  For example, 
SU(4) symmetry is of obvious importance to Gamow-Teller $\beta$-decay and decomposition into irreps could possibly help us understand, for example, 
unusually small matrix elements \cite{PhysRevLett.106.202502}. That will have to be left to future investigations.

\section[Nature vs. nurture]{Nature versus nurture, or: do random interactions naturally exhibit quasi-dynamical symmetries, or do they have to be coaxed?}

In my discussion above I excluded the IBM and the FDSM because they implicitly assumed dynamical symmetries. Certainly we know that 
nuclear spectra suggest symmetries, especially SU(3).  And textbooks typically lay the well-known fact that all even-even nuclides have ground states with 
angular momentum $J=0$ at the feet of the pairing force. 

The problem is that one can get similar results without those being good symmetries.  Rotational spectra, not just the energy levels but 
also ratios of reduced E2 transition matrix elements ($B$(E2)s), persist even when the actually SU(3) decomposition is badly fragmented, an observation which lead to the 
concept of quasi-dynamiical symmetries.  Worse, numerical experiments demonstrated that randomly chosen two-body interactions would tend to have $J=0$ ground states,
even if one sets all pairing-like matrix elements to zero \cite{PhysRevLett.80.2749,zelevinsky2004nuclear,zhao2004regularities}.

Detailed investigations suggest random interactions echo many of the rich phenomena of real nuclei.  Not only do random interactions tend to lead to $J=0$ ground states, 
one sees an odd-even staggering in binding energies along an `isotopic chain,' pairing gaps, and both particle-hole and pairing collectivity \cite{zelevinsky2004nuclear,zhao2004regularities,PhysRevC.61.014311}. The ratios of excitation energies 
for the first $J=0,2,4,6,8$ states are fairly tightly constrained to a single family spanning pairing-like, vibrational-like, and rotational-like spectra \cite{PhysRevC.75.047305}. 
Boson models with 
random interactions show a tight correlation between ratios of excitation energies and ratios of $B$(E2)s \cite{PhysRevLett.84.420}, a correlation not found in fermion models.  

These phenomena are not the work of true dynamical symmetries; for example, even with degenerate single-particle energies the $J=0$ ground states are not 
dominated by seniority zero (V. Zelevinsky, private communication).  Very little work has been expended on looking for 
quasi-dynamical symmetries in random interactions; 
to do one would need a robust way to compare decompositions into irreps in a fast, efficient, and above all appropriate manner.  It may be worthwhile to do so, 
as despite numerous papers written on the topic no general understanding has yet arisen \cite{PhysRevLett.100.162501}.

\section{Symmetry-guided bases}

\label{Symp}

One motivation for  decomposing wave functions into group irreps is the possibility of more compact bases for many-body calculations. By using a basis with more correlations, 
and especially of the \textit{right} correlations built in, one may describe nuclei with not billions but just tens of states.  Of course those correlated states are themselves complex, 
but if they belong to group irreps we know many of their properties in advance.

The main effort, in case you have not guessed, is in SU(3); the main barriers are, first, the high fragmentation across irreps, and, second, the difficulty of the group theory. 
The first has lead to the so-called pseudo-SU(3) basis \cite{arima1969pseudo,raju1973search,castanos1992transformation}, a discussion of which is beyond the scope 
of this paper. Because $j$-$j$ coupling is much easier to implement and to harness to geometrically growing computational power, the strongest application of SU(3) 
has been not in phenomenology but in \textit{ab initio} calculations. One barrier to standard no-core shell model calculations in the $j$-$j$ scheme is that the harmonic 
oscillator basis allow for exact projection of spurious center-of-mass motion but do poorly in describing the exponential tail of the wavefunction; brute force inclusion of 
many oscillator shells leads to a crippling explosion in dimensions. By contrast, the multi-shell generalization of SU(3), 
the symplectic group Sp(3,R), naturally couples to higher shells while still projecting spurious states, leading to better radial wave functions and description of, for example, 
electromagnetic interactions \cite{PhysRevLett.82.5221}. 

More recent work has demonstrated Sp(3,R) is a strong quasi-dynamical symmetry arising naturally in \textit{ab initio} calculations 
\cite{PhysRevLett.98.162503,PhysRevC.76.014315} . 
Success in applying this insight to structure calculations  \cite{0954-3899-35-9-095101,0954-3899-35-12-123101,1742-6596-569-1-012061} 
did not use symplectic symmetry 
directly, but instead used a tower of SU(3) irreps, built in different harmonic oscillator shells. Such work opens up the possibility of carrying out calculations not currently 
possible in the standard $j$-$j$ coupled NCSM, such as describing the Hoyle state in $^{12}$C, known to require many excitations in a harmonic oscillator 
space \cite{1742-6596-403-1-011001}. 
Although this approach is very promising, only time will tell for sure if the advantages gained by group theory will outweigh the technical difficulties needed to implement. 
This question is very much like that one I brought up at the beginning of this paper, the simplicity of the $M$-scheme against the compact correlations of the $J$-scheme.





\section{Summary and Acknowledgements}

I have given a very brief overview of the many rich applications of group-theoretic decompositions to nuclear structure. The most important points are that while 
group symmetries are often badly broken, leading to fragmented distribution over many group irreducible representations, 
we still see strong quasi-dynamical symmetry, that is, coherent 
structures persistent across many states. Quasi-dynamical symmetry is most evident in rotational bands, but we can see it in non-rotational bands as well, illustrated 
here by $^{48}$V.  Assuming pure dynamical symmetries, that is dominance by a single irrep, is an oversimplification, but the ubiquity and persistence of quasi-dynamical 
symmetry is a strong motivation for using symmetry-adapted bases for nuclear structure. That story is not over, as we still have to find the balance between elegance 
and practicality

I thank P. Navratil for use of his code to generate and SRG evolve the \textit{ab initio} chiral interaction. 
I would also like to thank D. J. Millener, J. Escher, and M. A. Caprio for useful conversations in recent years, especially regarding SU(3) and SU(4), though
any errors in describing group theory are mine alone. 
 This material is based upon work supported by the U.S. Department of Energy, Office of Science, Office of Nuclear Physics, under 
Award Number  DE-FG02-96ER40985.    



\bibliographystyle{apsrev4-1}
\bibliography{johnsonmaster}
\end{document}